\documentclass[aps,prd,twocolumn,floatfix,noshowpacs,tightenlines,noshowkeys,superscriptaddress,amsmath,amssymb,nofootinbib]{revtex4}
\usepackage{amssymb,amsbsy,epsfig,color,graphicx}
\usepackage{color}
\usepackage{[longtable}
\usepackage{array}
\usepackage{dcolumn}   
\usepackage{cellspace}
\usepackage{mathtools}
\usepackage{amstext}
\usepackage{amssymb}
\usepackage{stackrel}
\usepackage{graphicx}
\usepackage{esint}
\usepackage[utf8]{inputenc}
\usepackage{blindtext}
\usepackage{float}
\restylefloat{table}
\usepackage{booktabs}
\usepackage{enumitem} 

\usepackage{etoolbox} 
\usepackage{lipsum} 
\usepackage[capitalize]{cleveref}
\usepackage{multirow}
\usepackage[caption=false]{subfig}
\renewcommand\[{\begin{equation}}
\renewcommand\]{\end{equation}}

\newcommand{\ba}{\begin{eqnarray}}
\newcommand{\ea}{\end{eqnarray}}
\newcommand{\LF}{\left(}
\newcommand{\RF}{\right)}
\newcommand{\LT}{\left[}
\newcommand{\RT}{\right]}

\renewcommand\({\left(}
\renewcommand\){\right)}
\renewcommand\[{\left[}
\renewcommand\]{\right]}

\newcommand{\be}{\begin{equation}}
\newcommand{\ee}{\end{equation}}                  
\newcommand{\refsec}{Section~}

\def\beq{\begin{equation}}
\def\eeq{\end{equation}}

\makeatletter

\appto{\appendix}{%
\@ifstar{\def\theequation@prefix{A.}}%
{}%
}
\makeatother

\begin{document}

\title{Scalar Dark Matter Probes the Scale of Non-locality }

\author{Anish Ghoshal}
\affiliation{Dipartimento di Matematica e Fisica,  Universit\`a Roma Tre, 00146 Rome, Italy}
\affiliation{Laboratori Nazionale di Frascati-INFN, C.P. 13, 100044, Frascati, Italy}


\begin{abstract}
Scalar dark matter (DM) in a theory introduces hierarchy problems, and suffers from the inability to predict the preferred mass range for the DM.
In a WIMP-like minimal scalar DM set-up we show that the infinite derivative theory can predict DM mass and its coupling. The scale of non-locality (M) 
in such a theory in its lower-most limit (constrained by LHC) implies a DM mass $\sim$ TeV and a coupling with the Standard 
Model (SM) Higgs  $\lambda_\mathrm{HS} \sim 10^{-2}$. Planned DM direct detection experiments reaching such sensitivity in the DM will effectively
translate into lower bounds on the scale at which the non-locality comes into the play. 
\end{abstract}

\maketitle


\section{Introduction}
\label{sec:intro}
\medskip
Renormalization in quantum field theory deals to remove the ultraviolet (UV) divergences as in the case
of quantum electrodynamics. In the context to the Standard Model (SM), however, this leads to the well-known hierarchy
problem for the SM Higgs, which is basically the huge seventeen order of magnitude difference between the Planck and the electroweak (EW) scales \cite{Haber:1984rc}.
Solutions of this problem exist aplenty, most of which go by including a plethora of new particles at the or near the (EW) scale, including that of supersymmetry (SUSY)
where the bosonic and their fermionic loop corrections to the Higgs mass exactly cancel each other due to the presence of the super-partners.
With no hint of BSM particles whatsoever at the LHC or other experiments \footnote{None at $\geq 5 \sigma$ level at least. The $^8$Be anomaly \cite{Krasznahorkay:2015iga} is
an exception to this but it still needs to confirmed (see Ref. \cite{Nardi:2018cxi} and the references therein, for future prospects to confirm this).}, alternative solutions to this problem by other means have 
gained serious momentum in recent times, namely the relaxion \cite{Graham:2015cka}, Higgsplosion \cite{Khoze:2017tjt}, Clockwork \cite{Giudice:2016yja} and No-scale theories \cite{Salvio:2014soa}, to name a selected few.

Motivated by string field theory \cite{sft1,sft2,sft3,padic1,padic2,padic3,Frampton-padic,Tseytlin:1995uq,marc,Siegel:2003vt}, infinite derivatives provide a diligent approach to address this divergence problem by generalizing the kinetic energy operators of the Standard Model (SM) to
an infinite series of higher order derivatives suppressed by the scale of non-locality (M) at which the higher order derivatives come into play \cite{Biswas:2014yia}.
Furthermore, the negative running of the self-interacting term for the SM Higgs which gives to rise to a metastable vacuum \cite{Olive:2016xmw} was also cured by investigating
the RGE of the theory \cite{Ghoshal:2017egr}. It was discovered that the $\beta$-functions at the scale of non-locality, and the wavefunction renormalization meant the fields
are frozen beyond M. To be precise, capturing the infinite derivatives by exponential of an entire function softened UV behaviour in the desirable manner without 
introducing any new degrees of freedom in the particle spectrum, as they contain no new poles in the propagators. They have been explicitly shown to be ghost-free 
\cite{Buoninfante:2018mre} and provides unique scattering phenomenology rendering transmutation of energy scale which has its own cosmological implications \cite{Buoninfante:2018gce}.

On the gravity side, Ref.\cite{Biswas:2011ar} showed the most general quadratic curvature gravitational action (parity-invariant and 
torsion-free), with infinite covariant derivatives can make the gravitational sector free from the Weyl {\it ghost} and, is free
from classical singularities, such as blackhole 
ones~\cite{Biswas:2011ar,Biswas:2013cha,Frolov:2015bia,Frolov:2015usa,Koshelev:2018hpt,Koshelev:2017bxd,Buoninfante:2018xiw,Cornell:2017irh,Buoninfante:2018rlq,
Buoninfante:2018stt}~\footnote{Previously, arguments were provided regarding non-singular solutions in Refs.~\cite{Tseytlin:1995uq,Siegel:2003vt}.} and cosmological 
ones~\cite{Biswas:2005qr,Biswas:2006bs,Biswas:2010zk,Biswas:2012bp,Koshelev:2012qn,Koshelev:2018rau}. 

Now on one hand, it remains to build a non-Abelian aspects of the theory \cite{Anish2} and on the other investigate if the theory may provide solutions to some other
problems that engulf the SM currently. The non-baryonic matter component of the universe, the non-luminous dark matter (DM), constitutes about
a fifth of the total energy density of our universe -- fact now well established courtsey to several
evidences in cosmology and astrophysics at different scales \cite{Bertone:2010zza}. The observation of DM in the universe has no explanation in the SM. Despite
the several experimental searches and the impressive efforts of the community, all the evidence for DM has only been of gravitational nature and 
the widely considered non-gravitational nature of DM is still unknown \cite{Jungman:1995df}. There are many candidates postulated
for this \cite{Bertone:2004pz, Mazumdar:2011zd}, and a lot
of attention was devoted to the class of beyond Standard
Model (SM) theories which can provide a weakly interacting
massive particle (WIMP) DM candidate, which, as the name suggests, weakly interact with SM particles and
contains an appealing connection between the dark sector and the electroweak scale  \cite{Arcadi:2017kky}. 
Colorless, electrically neutral
and weakly interacting massive particles with mass in the GeV-TeV range are ubiquitous in new physics
models, and appear to be well suited to reproduce quantitatively the measured DM energy density if their
stability on cosmological time scales can be ensured.
Moreover, WIMP DM particles were in thermal
equilibrium with the SM in the early Universe; and thus, were produced via the standard
freeze-out mechanism. This provided several avenues to search
for DM: direct detection \cite{Akerib:2016vxi, Cui:2017nnn, Aprile:2018dbl} and indirect detection \cite{Aguilar:2016kjl, Fermi-LAT:2016uux}, as well as the
collider searches for DM \cite{Aaboud:2016obm,Sirunyan:2017hnk}. But none of them have found any such interactions hitherto. The most
constraining limits for WIMP DM come from the direct detection experiments, especially for DM masses in the GeV ballpark.

In particular we consider WIMP-like scalar DM models and follow the same idea as in Ref. \cite{Khoze:2018bwa} which
relates the properties of scalar DM to the Higgsplosion scale. Any fundamental scalar has the same naturalness problem like that of the SM Higgs, described earlier. 
Thus the situation renders the DM mass (or DM mass range) to be unpredictable.
Here we show that infinite derivative model not only solves this problem but also provides a relation between observation (DM reslic density), DM mass, its coupling 
to SM and the non-local scale (M). Thus the DM phenomenology is dictated at the non-local scale. \footnote{From the non-local gravity side, there exists a 
treatment of DM in Ref. \cite{Boos:2018bxf,Hehl:2008eu}.} Furthermore, M which is constrained by collider
experiments \footnote{The authors of Ref. \cite{Biswas:2014yia} had only roughly estimated for the production processes at the LHC.
Here, on the contrary we perform explicit cross-section calculations and put impeccable bounds on M.} maybe probed from existing and planned DM direct 
and indirect detection experiments \cite{Biswas:2014yia}. 
Such probes of the parameter space regions are far better than that from the colliders, going upto 30 TeV or so.

The paper is organized as follows: we describe scalar infinite derivative model in section 1. In the next section we study the freeze-out mechanism of the DM. This is followed
direct detection section. In the final section, we conclude by discussing some of the impactful aspects of our study.
\medskip

\section{Infinite Derivative Scalar Theory}
The action for the infinite derivative theory is given by \cite{Biswas:2014yia}:
\be \label{Action}
S = \int d^4 x\ \LT-\frac{1}{2} \phi e^{{\Box+m^2\over M^2}}(\Box + m^2)\phi -{\lambda \over 4!}\phi^{4}\RT
\ee
Here  the normalization of $\phi$ is so chosen that the residue at the $p^2=m^2$ pole is unity.
$\Box= \eta_{\mu\nu}\partial^{\mu}\partial^{\nu}$ $(\mu, \nu=0,1,2,3$) with the convention of the metric signature $(+,-,-,-)$, 
  $m_{\phi}$ are the masses of the scalar, 
  and $M$ is the scale of the non-locality which is taken to be below Planck scale. 
In our non-local field theory, the kinetic terms are generalized with higher derivatives suppressed by non-local scale $M$, 
 while the scalar self-interaction is the standard one. 
The theory is reduced into the standard local field theory in the limit of $M \to \infty$.
In Eucleadian space ($p^0\rightarrow ip^0$) the propagator is given by,
\be \label{Prop}
\Pi(p^2)=-{ie^{-{p^2+m^2\over M^2}}\over p^2+m^2}
\ee
while the vertex factor is, as usual, given by $-i\lambda$.
Note that the non-local extension of the theory leads to the exponential suppression of the propagators for $p_E^2 > M^2$, 
and this fact indicates that quantum corrections will be frozen at energies higher than $M$. 

Although the action in Eqn. \ref{Action} shows a  modification in the kinetic term, 
however, note that an equivalent description can be done with the usual local Klein-Gordon kinetic operator by making 
the following field re-definition \cite{Buoninfante:2018mre}:
\begin{equation}
\begin{array}{rl}
\tilde{\phi}(x)= & \displaystyle e^{-\frac{1}{2}(\Box)}\phi(x)\\
= & \displaystyle \int d^4y \mathcal{F}(x-y)\phi(y),
\end{array}
\label{42}
\end{equation}
where $\mathcal{F}(x-y):=e^{-\frac{1}{2}(\Box)}\delta^{(4)}(x-y)$. 
However, the real fields of the theory are $\phi$, and not $\tilde{\phi}$. With the field redefinition into the action becomes:
\begin{equation}
S=\frac{1}{2}\int d^4x \tilde{\phi}(x)(\Box)\tilde{\phi}(x) -\int d^4x \frac{e^{\frac{1}{2}(\Box)} g}{4!}\tilde{\phi}(x).
\label{43}
\end{equation}
Eq.\eqref{43} shows the form factor $e^{\frac{1}{2}(\Box)}$ appears in the interaction term, thereby non-locality is significant only when 
the interaction is switched on as the free-part remains just the standard local Klein-Gordon kinetic term. So, this can be cosidered as an
interaction-level theory only.

We use the former description using the $\phi$ field definition throughout the paper. 
For momenta k and p,
$\bar{k} = \frac{k}{M}$ and $\bar{p} = \frac{p}{M}$ notations are used.
\medskip

\section{Scalar Particle Dark Matter }
\label{sec:pheno}

\subsection{Model}
\label{sec:model}

We take the simplest possible extension to the SM of particle physics, precisely, a singlet real scalar, $S$, with a $\mathbb{Z}_2$ symmetry (well studied in the 
literature, see Refs. \cite{McDonald:1993ex,Arcadi:2017kky} for a selected few). The $\mathbb{Z}_2$ symmetry makes the DM stable, and maybe a bi-product of breaking
chains from higher GUT groups.

For this simple scenario, SM sector and the DM sectors are in contact only via a Higgs-portal coupling, $\lambda_{H S}$,
\begin{equation}
\mathcal{L}_{BSM} = \frac{1}{2}\partial_\mu S \partial^\mu S - \frac{1}{2}m_{S}^2 S^2 - \frac{\lambda_{S}}{4!} S^4 - \frac{\lambda_{H S}}{2} S^2 \left(H^\dagger H\right) ,
\end{equation}
with,
\begin{equation}
\mathcal{L}_{SM} \supset V(H) \sim m_{H}^2 H^\dagger H - \lambda_{H}(H^\dagger H)^2.
\end{equation}
We generalize the kinetic terms and introduce M as the scale of non-locality at which the higher derivatives come into play. Thus 
\begin{equation}
\frac{1}{2}\partial_{\mu} S \partial^{\mu} S \rightarrow \frac{1}{2} \partial_{\mu} S e^{\frac{\Box + m ^2 _{S} }{M ^2}} \partial^{\mu} S
\end{equation}
Coonsidering $\lambda_{S}\ll\lambda_{H S}$ will be in the lines of \refsec {RGE} where we present the renormalization group equation (RGE) flow of the masses and the coupling. 
Bare mass term in the dark sector is ignored in order to not introduce any other energy scale in the model.
the BSM scalar mass when corrected with RGE contributions, consists of the quadratically divergent contribution from loops of the SM
Higgs particle. In case for the standard local field theory, the renormalised mass is given by:
\begin{equation}
	m_{S}^2 = m_{S,0}^2+\delta m_{S}^2\approx\delta m_{S}^2\approx \frac{\lambda_{H S} \Lambda_{\rm{UV}}^2}{16\pi^2},
\end{equation}
where $\Lambda_{\rm{UV}}$ is the UV cut-off of the theory. 
In the infinite derivative scalar theory, as discussed in Ref. \cite{Biswas:2014yia}, the mass corrections will be dominated by the Gaussian correction coming from the 
non-local propagator \ref{Prop}. The renormalization cut-off is governed by M basically and in the UV, the theory remains finite.
The mass correction is given by:
\begin{widetext}
\be
\delta m_{s} ^2=i\Gamma_2=-{i\lambda_{HS} \over 2}\int {d^4k\over (2\pi)^4}\ {e^{-{(k^2+m_{s} ^2)\over M^2}}\over k^2+m_{s} ^2}
={\lambda_{HS} \over 32\pi^2}\left[e^{-{m_{s} ^2\over M^2}} +\LF{m_{s} ^2\over M^2}\RF  Ei \left( -\frac{m_{s} ^2}{M^2} \right)  \right]M^2
\label{deltam2}
\ee
\end{widetext}
where $Ei(x)$ is the exponential-integral function defined by:
\begin{equation}
 Ei(x) = \int _{-\infty} ^{x} \frac{e^{-t}}{t} 
\end{equation}
The $Ei$ function has a mild divergence as $z\rightarrow 0$, but $z Ei(z)\rightarrow 0$ as $z\rightarrow 0$. Thus we see that when $M\gg m$
\be
\delta m_{S} ^2={\lambda \over 32\pi^2}M^2
\label{deltamsquare}
\ee
Therefore, in this region of the parameter space, the BSM scalar has mass
\begin{equation}
m_{S}^2 \approx \lambda_{H S}\frac{M^2}{16\pi^2}.
\label{eq:massscale}
\end{equation}
If the non-local scale is known, the the parameter space is bounded from this particular relation.

\subsection{RGE -- Mass}
In SM, the Higgs bare-mass needs to be fine-tuned in the UV so that the measured Higgs mass is in agreement in the electroweak scale  \cite{Aad:2012tfa,
Chatrchyan:2012xdj}.   
In case of the infinite derivative theory, the Higgs bare mass depends on \cite{Biswas:2014yia}:
\begin{equation}
m_{H} ^2 = -\frac{\lambda_{H}}{2} v_{EW}^2 - M^2 e^{-\frac{m_{S} ^2}{M^2}} \( \frac{\lambda_{H}}{4 \pi^2}  +\frac{\lambda_{H S}}{16 \pi^2} - \frac{N_c y_t^2}{8\pi^2} + \cdots \), 
\end{equation}
where the first term is the Higgs mass and the rest of the terms belong to quantum self-energy corrections, with $N_c$ being the no. of colors of quarks.
In the limit $M\rightarrow \infty$, we recover the SM $\beta_m$, and for scales $\geq$ M, the $\beta_m$ function vanishes thereby rendering the theory scale invarianct \cite{Anish2}.
Here, one can say the Higgs mass is fine-tuned to the extent of $\sqrt{\lambda_{H}}v_{\rm{EW}}/ M$, 
which in comparison to fine-tuning of order $m_H/m_{\rm GUT} \sim 10^{-14}$ is pretty less fined-tuned, for M$\leq M_{GUT}$.

In order to keep the self-energy contribution dominant we assume, the bare S mass,
\begin{equation}
- \frac{\lambda_{H S} e^{\frac{-m_S^2}{M^2}} }{16\pi^2}M^2 \lesssim m_{S,0}^2 \lesssim \frac{\lambda_{H S} e^{\frac{-m_S^2}{M^2}}}{16\pi^2}M^2.
\end{equation}

\subsection{RGE -- Quartic and a Stable Vacuum}
In SM, the Higgs quartic runs to become negative at $\sim 10^{11-12}$ GeV energy scales, depending on the top quark mass uncertainty. Consequently this makes
the vacuum metastable. Such a vacuum is not compatible with the observable universe and especially when inflation is considered, which nonetheless
would drive the vacuum to instability due to evolving Higgs fluctuations \cite{East:2016anr}. This being a serious problem, was studied in the infinite derivative  context and
it was shown that Higgs quartic does not run to be negative in the UV \cite{Anish2}.

\section{Dark Matter formation by Freeze-Out}
\label{sec:freeze}

DM if thermally produced in the early the procudure of DM formation follows the standard freeze-out scenario.
Following Ref. \cite{Burgess:2000yq}, the thermally-averaged dark matter annihilation cross-section $\langle \sigma_{ann}v_{rel}\rangle \equiv \langle \sigma v\rangle$, 
in the region of the paprameter space $m_{S} \gg v$, where $v$ is the Standard Model Higgs vacuum expectation value, 
and that couple to $S$ through the $\lambda_{H S} S^2 H^\dagger H$ operator. The DM annihilation cross-section in this case is given by:
\begin{equation}
\langle\sigma v\rangle\approx \frac{\lambda^2_{HS} e^{\frac{-4 m_S ^2}{M^2}} }{16\pi m_{S} ^2 } .
\label{eq:approx}
\end{equation}
the exponential factor coming from the $\lambda$ RGE \& the croos-section calculation \cite{Biswas:2014yia,Ghoshal:2017egr}.
For $m_S \lesssim v$ values for the M scale that are too, low therefore the low-mass regime is not considered. 
Subsequently, the approximation of Eqn. \ref{eq:approx} will be considered henceforth.

The standard Boltzmann Equation for DM relic will be slightly modified due to the
presence of the infinite derivatives in the scalar kinetic operators but the final exact answer will differ only by a factor in the exponential thus giving us an opportunity to
present an approximate result using our present considerations. A detailed Boltmann solution including relic, direct, indirect and LHC production of scalar, fermionic and vector
DM scenarios is being considered in another publication \cite{Anish3}.
The relic density of dark matter today is constrained by the Planck satellite  \cite{Aghanim:2018eyx} to $\Omega_{\rm DM}h^2 = 0.12$.  
Our prediction for the present day density of $S$ particles is \cite{Khoze:2018bwa}:
\begin{equation}
\Omega_\mathrm{X}h^2 = \left[  \frac{8\pi G g_*(m_{S}/x_f)}{45} \right]^{1/2} \frac{ 4\pi^2 G x_f T_0^3}{45\langle \sigma v \rangle  H_{100}^2},
\label{eq:freezeout}
\end{equation} 
where $T_0$ is the current temperature, $g_*(m_{S}/fx_f)$ is the number of degrees of freedom in equilibrium during annihilation, $H_{100}=100$km s$^{-1}$ Mpc$^{-1}$, 
and $G$ the Newton's gravitational constant. 
We note $x_f=m_{S}/T_F$, also known as the inverse freeze-out temperature is given as logarithmic functions of the thermally averaged cross-section.

Solving the above equation for a perturbative range of $\lambda_{H S}$ yields the usual $x_f$ values in the range 20-30, 
the freeze-out condition in Eqn. (\ref{eq:freezeout}) gives a salient relationship between $\lambda_{H S}$, $m_{S}$ \& M as:
\begin{equation}
\lambda_{H S} e^{\frac{- 2 m_S ^2}{M^2}}= 0.30 \(\frac{x_f}{20}\)^{1/2}\(\frac{\Omega_\mathrm{X}h^2}{0.12}\)^{1/2}\(\frac{m_{S}}{1\mathrm{ TeV}}\).
\end{equation}

Thus we see that the combinations of a $\lambda_{H S}$ coupling perturbative (-4 $\pi \leq \lambda_{H S}  \leq 4 \pi$) 
and ensuring the correct relic density greatly limits the DM mass range which maybe allowed in the non-local infinite derivative scenario. 

We do not consider any fine-tuning between the bare mass term and the $\lambda_{H S}$ loop contribution.

\medskip

\section{Direct detection of DM}
\label{sec:direct}
In this section we consider the phenomenology involving DM crossing the earth being scattered by nucleons in direct detection experiments. These consist of
directly measuring their scattering off a detector’s target material, for massive particles producing recoil energies in the keV energy scale.

\begin{figure}[H]
\begin{center}
\includegraphics[width=7cm, height=6cm]{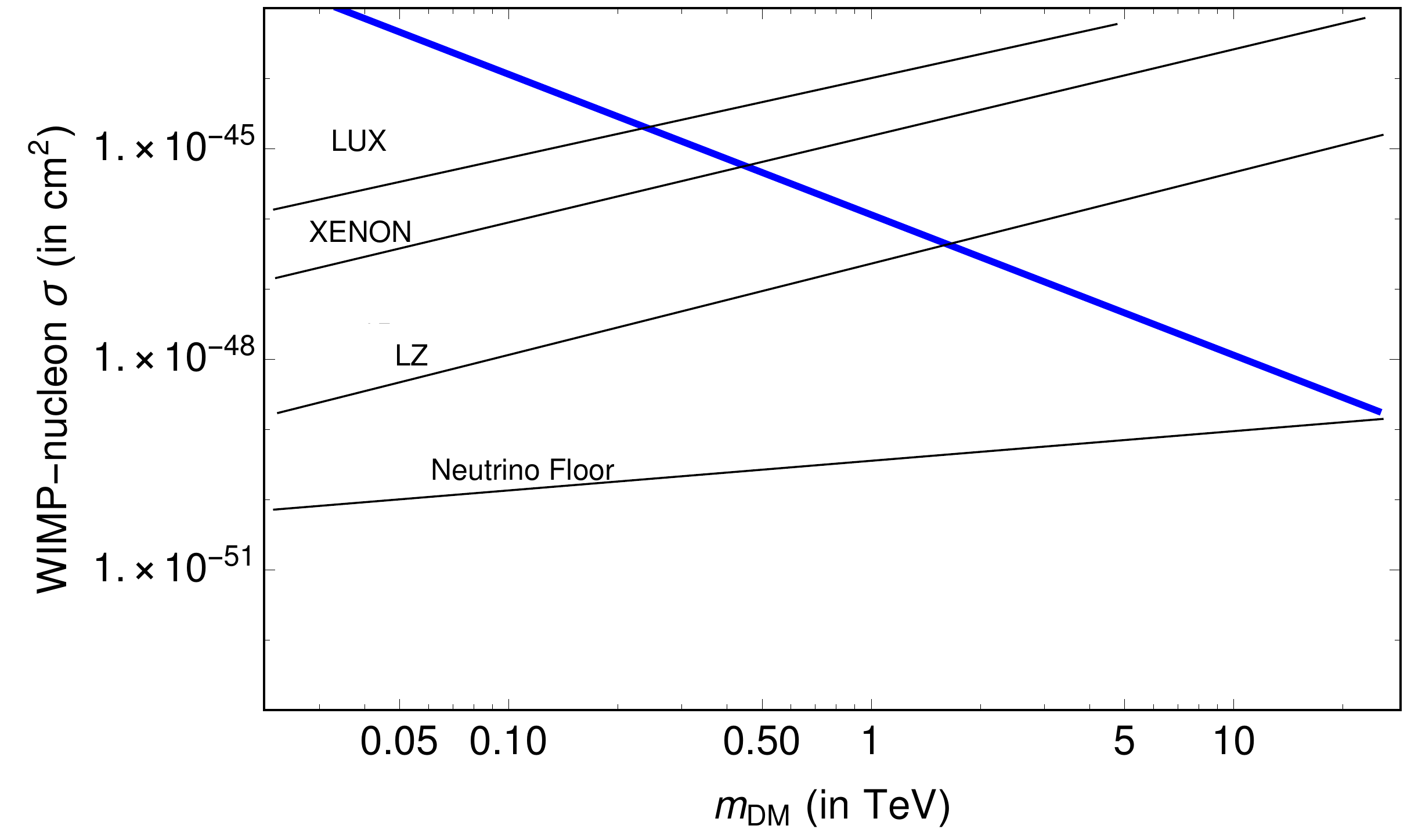}
\caption{Direct Detection of DM, with M $= 10^6$ GeV, $\lambda_{HS} = 10^{-1}$ chosen. Region above the LUX, XENON and LZ lines are/will be probed by the experiments.}
\label{Fig1}
\end{center}
\end{figure}

\begin{figure}[H]
\begin{center}
\includegraphics[width=7cm, height=6cm]{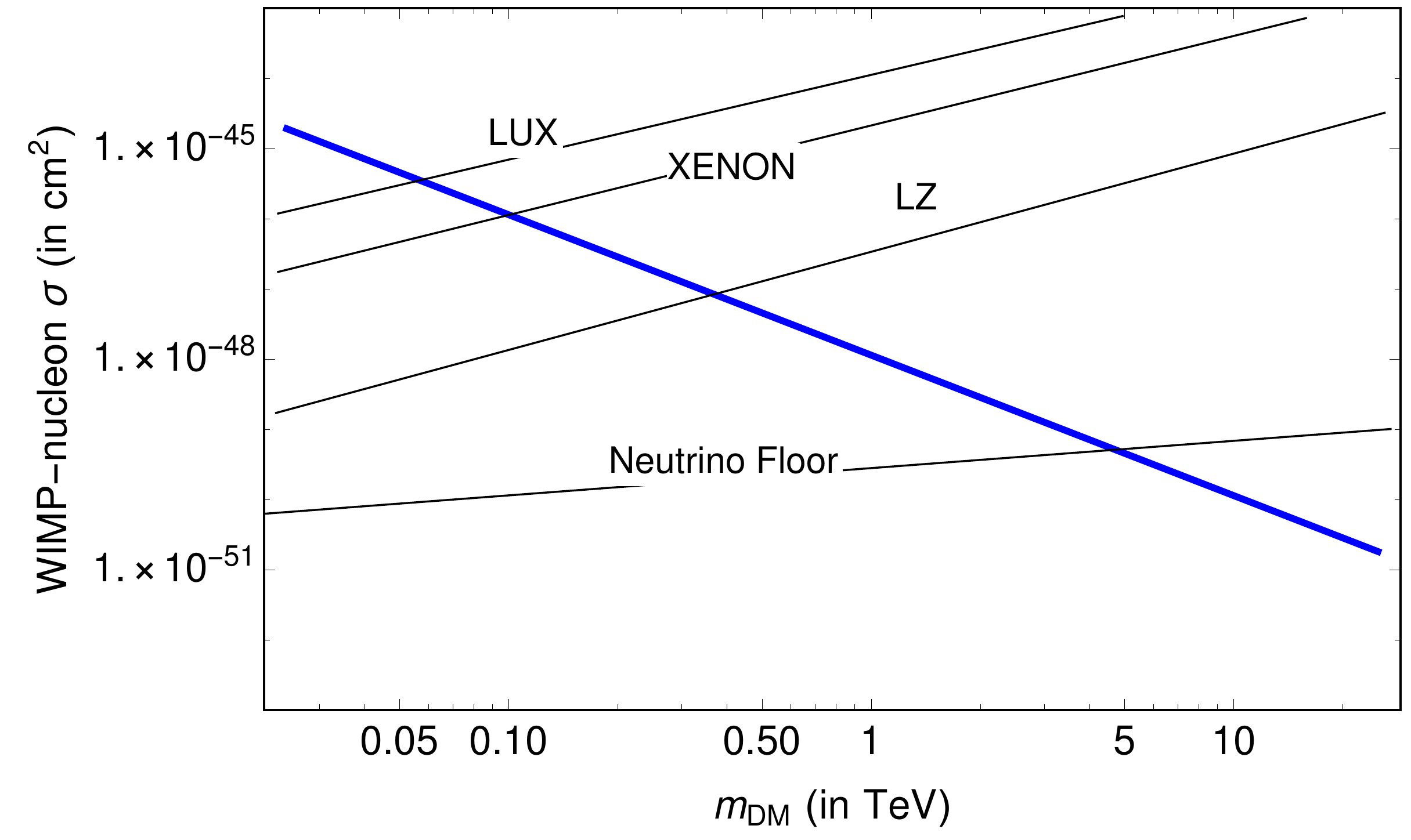}
\caption{Direct Detection of DM, with M $= 10^6$ GeV, $\lambda_{HS} = 10^{-2}$ chosen. Region above the LUX, XENON and LZ lines are/will be probed by the experiments.}
\label{Fig2}
\end{center}
\end{figure}

When estimating the cross-sections in direct detection experiments, values of $m_{S}$ and $\lambda_{H S}$ that are allowed by the relic density are kept in mind, while we consider 
DM-nucleon cross-sections as in \cite{Cline:2013gha, Khoze:2017tjt}:
\begin{widetext}
\begin{equation}
\sigma_{SI}= \frac{\lambda_{H S}^2 e^{\frac{-4 m_S^2}{M^2}} f_N^2}{4\pi} \frac{m_N^2 m_{S}^2}{(m_N + m_{S})^2} \frac{m_N^2}{m_H^4 m_{S}^2} \sim  \frac{\lambda_{H S}^2  e^{\frac{-4 m_S^2}{M^2}} f_N^2}{4\pi} \frac{m_N^4 }{m_H^4 m_{S}^2}, 
\label{eq:sigSI}
\end{equation}
where $m_N$ is the mass of a nucleon and $f_N \sim 0.3$ \cite{Alarcon:2011zs,Alarcon:2012nr} is an effective Higgs-nucleon-nucleon coupling.
\end{widetext}

\begin{figure}[H]
\begin{center}
\includegraphics[width=7cm, height=6cm]{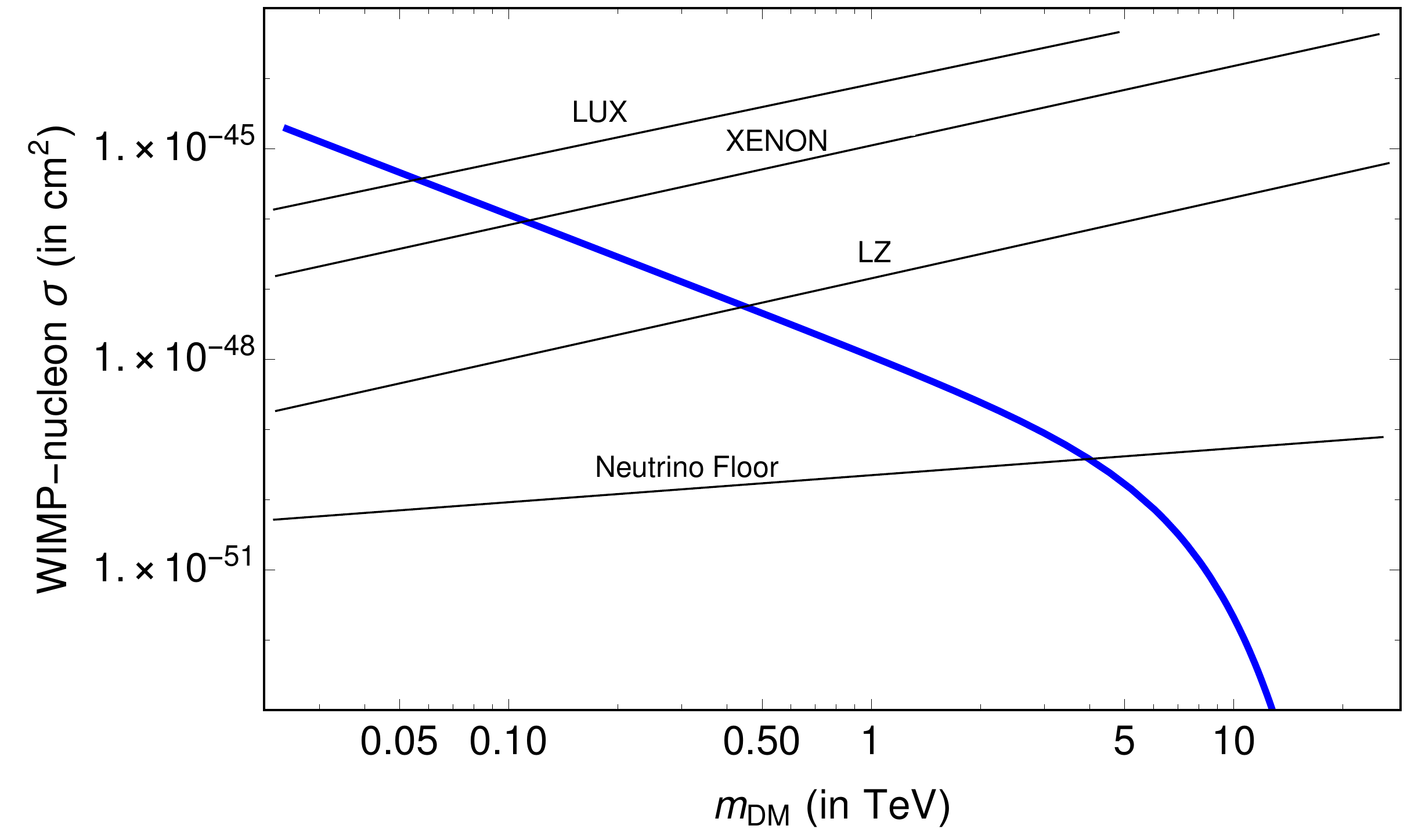}
\caption{Direct Detection of DM, with M $= 10^4$ GeV, $\lambda_{HS} = 10^{-2}$ chosen. Region above the LUX, XENON and LZ lines are/will be probed by the experiments.}
\label{Fig3}
\end{center}
\end{figure}

\begin{figure}[H]
\begin{center}
\includegraphics[width=7cm, height=6cm]{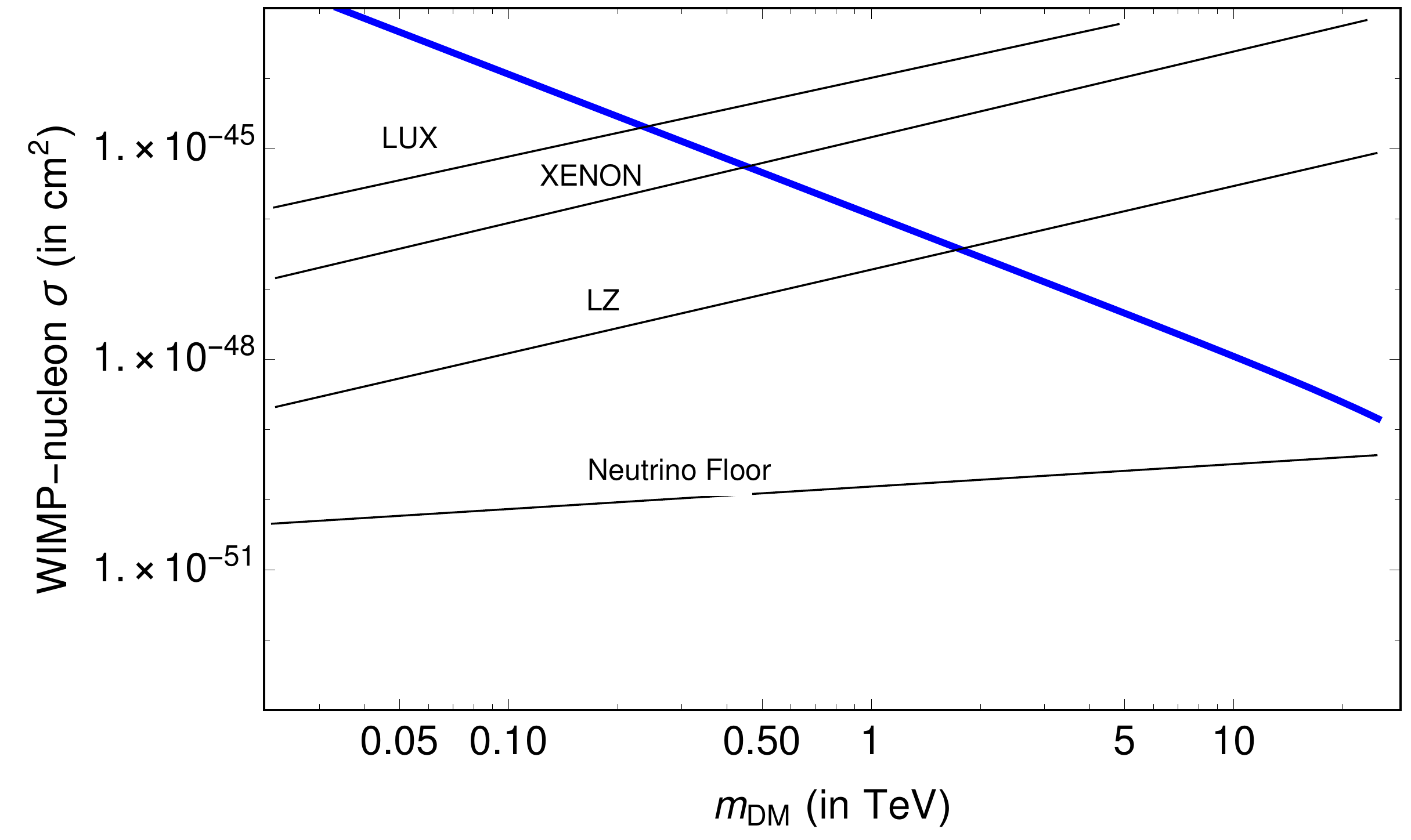}
\caption{Direct Detection of DM, with M $= 10^5$ GeV, $\lambda_{HS} = 10^{-1}$ chosen. Region above the LUX, XENON and LZ lines are/will be probed by the experiments.}
\label{Fig4}
\end{center}
\end{figure}

The DM mass m$_S$ gives us $\lambda_{H S}$ for which Planck relic density is satisfied in case of a given non-local scale M, and so we can see how the direct detection 
elastic cross-section, $\sigma_{SI}$, varies $m_{S}$ ( plotted in Figs. (1-4)). We have varied the non-local scale M,
and the associated quartic coupling, $\lambda_{H S}$; the blue line indicates the points which suit the values. Constraints from 
LUX \cite{Akerib:2016vxi}, Xenon-1T \cite{Aprile:2017iyp} and LZ \cite{Mount:2017qzi} are shown for reference, which are current and are projected bounds.
 We see that current constraints exclude DM masses below $\sim 2$ TeV for M at $10^6$ GeV. Future searches for dark matter such as LZ will probe rest of the
parameter space further and further.

\medskip

\section{Discussion and conclusions}
\label{sec:concl}

We studied an introductory version of a scalar DM theory and found it presents interesting phenomenology on one hand and predicts specific DM mass which is otherwise not
possible in standard theory. To list, we may have the following conclusions:
\begin{enumerate}
 \item The non-local infinite derivative theory results in a definite prediction for mass of a real scalar dark matter candidate, 
as well as a definite prediction of its coupling to the Standard Model fields. In particular, the lowest value for M scale theoretically preferred, i.e. $M \sim O(10)$ 
TeV, implies a dark matter mass of $m_{S} \sim 15$ TeV and a Higgs portal coupling $\lambda_{H S} \sim 0.5$, which remains safe from all current particle physics 
and cosmological constraints.
\item Even WIMP-like candidates with large coupling may form thermal DM. The observed small relic is due to the suppression due to presence the non-local scale $M_{NL}$ and is used to 
constrain it. 
\item DM direct detection experiments are probing the scale of non-locality ($M_{NL}$) in the O(10) TeV range. 
This is better than the current LHC reach as shown in Ref. \cite{Biswas:2014yia}.   
\end{enumerate}

Though we show this in a simple abelian set-up the prediction for DM direct detection in infinite-derivative field theory,
one may look to extend this to non-Abelian processes and perform the calculations for other dark matter portals. Including
psedoscalar or psedo-vector portals will give us spin-dependent cross-sections as well. However, in all these cases, our 
conclusion that the direct detection phenomenology probes the scale of non-locality much better than the standard LHC production
processes. 
We will show infinite derivative is useful to rescue the Higgs and fermion portal DM models in the regions of the parameter space,
otherwise heavily constrained, or even ruled out in standard theory in Ref. \cite{Anish3}. Production of DM in infinite derivative
theory is beyond the scope of our current study and will be taken up in future. To end we will also like to comment on other kinds of DM in the
infinite derivative context. This can be treated as a general structure of the infinite derivative DM theory: DM with large coupling to SM
sector today may not be in contradiction with current direct detection searches (in the ruled out parameter space regions) as the running in 
the theory dictates small coupling during the DM formation
which occurs at a higher energy scale ($\sim$ TeV) than that of the direct detection experiment ($\sim$ 200 MeV) which is a low-energy scattering phenomenon.

\bigskip

\section*{Acknowledgements}
The author thanks Anupam Mazumdar for discussions and comments to improve the work. Besides, he also thanks the
phD grant of LNF-INFN and Universita Roma Tre, along with its logistics support. 
He is grateful to Enrico Nardi for learning order-of-magnitude estimations beside some
discussions on cosmology. 


\newpage


\end{document}